# Role of wave-particle resonance in turbulent transport in toroidal plasmas

G. Dong[1] and Z. Lin[2,*]
[1]Princeton Plasma Physics Laboratory, Princeton University, Princeton, New Jersey 08540, USA
[2]Department of Physics and Astronomy, University of California, Irvine, California 92697, USA
*Email: zhihongl@uci.edu

**Keywords:** turbulence, transport, tokamak, simulation

**Abstract**

Wave-particle interaction in toroidal plasmas is an essential transport mechanism in drift wave instability-driven microturbulence. In tokamkas, different wave-particle resonance conditions have been found important for the energy and particle transport of multiple species in various drift wave turbulences. To confirm the transport mechanism for electrons and ions in tokamak drift-wave instabilities, the effect of wave-particle resonance on turbulent transport is studied using global gyrokinetic particle simulations of the plasma core ion temperature gradient (ITG) and collisionless trapped electron mode (CTEM) turbulence. Simulation results show that in CTEM and ITG turbulence, electron transport is primarily regulated by wave-particle linear resonance, and the ion transport is regulated by nonlinear wave-particle decorrelation.

## I. Introduction

Wave-particle interaction in toroidal plasmas is an essential transport mechanism in drift wave instability-driven microturbulence, which is a prominent candidate for the anomalous heat loss in magnetic fusion devices [1, 2]. In tokamkas, different wave-particle resonance conditions have been found important both theoretically and in simulations for the transport of thermal ions [3], electrons [4] and energetic particles [5] in various drift wave turbulences including ion temperature gradient (ITG) turbulence [3], electron temperature gradient (ETG) turbulence [4], and collisionless trapped electron mode (CTEM) turbulence [6].

In this work, the effect of wave-particle resonance on turbulent transport in fusion plasmas is studied using global gyrokinetic particle simulations of the ITG and CTEM turbulence. In the wave-dominated turbulence, the particle and heat transports are induced by the breaking of the second and the third adiabatic invariants in the tokamak due to the time variations of the fluctuating fields (linear wave-particle resonance) and the *EXB* drifts (nonlinear particle scatterings) [7]. To study the relative contributions of the linear and nonlinear effects, we compare the transport between dynamic plasma turbulence in fully self-consistent simulations and turbulence in test-particle simulations with suppressed wave-particle resonance. In the simulation, the contribution of wave-particle resonance can be eliminated by preventing the electromagnetic field to evolve with time. At the nonlinearly saturated stage, with fixed field and wave-particle resonance suppressed, thermal ion and electron transport rates can be calculated and compared to those in the nonlinear stage in standard ITG and CTEM turbulence with self-consistently evolving field. Simulation results show that in CTEM turbulence, when wave-particle resonance is suppressed, electron heat and particle transport decreased to a very low level, while ion heat transport level saw no dramatic change. Ion heat transport in ITG turbulence with suppressed wave-particle resonance also showed small qualitative difference from that in standard ITG turbulence.

## II. ITG turbulence

In this work, the gyrokinetic toroidal code (GTC) [6, 8] is utilized to study the linear and nonlinear wave-particle interactions in ITG and CTEM turbulences. First-principles gyrokinetic simulations of electrostatic ITG and CTEM turbulence are carried out using GTC which has extensively been applied to study turbulent transport including ITG [8], CTEM [9], and ETG [4] turbulences, energetic particle turbulent transport [5, 10] and kinetic MHD turbulences [11] in core and edge plasmas [12].

To study the ITG turbulence, the global GTC simulations use representative tokamak plasmas with concentric flux-surfaces and the following local parameters at a radial position *r = 0.5a, R/$L_{Ti}$ = 6.9, R/$L_n$ = 2.2, q =1.4, s=0.78, $T_e$/$T_i$ = 1 and a/R = 0.36*. Here *R* and *a* are the major and minor radii, $L_{Ti}$ and $L_n$ are the ion temperature and density gradient scale lengths, $T_i$ and $T_e$ are the ion and electron temperatures, respectively, *q* is the safety factor and *s* is the magnetic shear. The profile for the safety factor is *q = 0.581 + 1.092(r/a) + 1.092(r/a)$^2$* and for the temperature and density gradients is *exp{−[(r −0.5a)/0.32a]$^6$}*. The boundary condition of the perturbed electrostatic potential *ϕ = 0* is enforced at *r < 0.1a* and *r > 0.9a*. The size of the tokamak used in the simulation is *a = 250$\rho_i$*, where $\rho_i = v_i/\Omega_i$ is the ion gyroradius, ion thermal speed $v_i = (T_i/m_i)^{1/2}$ and ion mass $m_i$. The computational mesh consists of *64* toroidal grids and a set of unstructured radial and poloidal grids with a perpendicular grid size of $\rho_i$. The time step is *0.2$L_{Te}$/$v_i$*. In the ITG simulations, ions are governed by the gyrokinetic equation [13, 14] and electrons are assumed to be adiabatic

In the self-consistent nonlinear electrostatic GTC simulations using the plasma parameters described above, the ITG is the dominant instability. The time evolution of the ITG-induced transport is shown in Fig. 1 in the solid blue line. The self-consistent heat flux $q = \int dv^3 (\frac{1}{2}mv^2 - \frac{3}{2}T)\delta v_{E\times B}\delta f$ is used to define the effective ion heat conductivity $\chi$ by using the relation $q = n_0\chi\nabla T$, where v is the particle velocity, $\delta v_{E\times B}$ is the radial component of the gyroaveraged EXB drift, and $\delta f$ is the perturbed particle distribution function. The ITG instability first grows exponentially in the linear regime until around t=70 $R_0/v_i$, and then saturates due to the self-regulation by zonal flows. Finally, the turbulence is fully developed and the ion heat conductivity reaches an amplitude on the order of the gyro-Bohm level (*$\chi_{GB}=\rho_i^2 v_i/a$*). The particle flux is zero since electron is adiabatic.

In the wave-dominated turbulence, the ion transport is induced by the breaking of the second and the third adiabatic invariants in the tokamak due to the time variations of the fluctuating fields (linear resonance) and the *EXB* drifts (nonlinear scatterings). To delineate the relative contributions of the linear and nonlinear effects, we compare the transport between fully self-consistent simulation and test-particle simulation with fixed turbulence fields. In the test-particle simulation, we record the turbulence fields at *t=110 $R_0$/$v_i$* from the fully self-consistent simulation and restart the simulation with the fixed turbulence fields at *t=110 $R_0$/$v_i$*. The resulting ion heat conductivity is shown in black solid line in Figure 1. To prevent unphysical ion particle flux (charge separation), zonal flows are kept self-consistent, which induce the geodesic acoustic mode (GAM) oscillation with a frequency of *ω ~2.3$v_i$/$R_0$*. The observed GAM frequency is in agreement with the earlier theory [15] and simulation [16]. We find that the ion heat conductivity from the fixed fields changes quantitatively (within a factor of two) compared with that from the self-consistent dynamic fields, as shown in Figure 1.

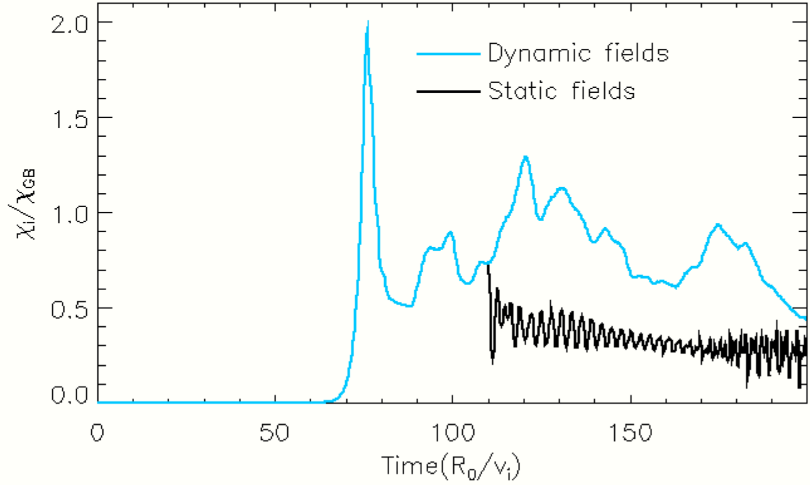

**Figure 1.** Time history of ion heat conductivity $\chi_i$ with self-consistent dynamic fields (blue) and fixed fields (black) in the nonlinear regime. The ion heat conductivity from the fixed fields changes within a factor of two compared with that from the self-consistent fields.

To verify that the test particle transport is induced by the *EXB* scatterings, we examine the phase space structure of $\chi_i$. Since ion transport is diffusive in the ITG turbulence, the phase space structure of the ion heat conductivity can be calculated accurately through the ion mobility $D_{eff}$, defined as:

$$D_{eff} = \frac{1}{2N\Delta t}\sum_{i=1}^{N}\Delta r_i^2 \quad (1)$$

where $\Delta r_i = r_i(t+\Delta t) - r_i(t)$ is the radial displacement of the guiding centers, and $i=[1,N]$ denotes the particle label. Equation (**1**) is valid for any time separation $\Delta t$ that is longer than the effective wave-particle decorrelation time. Time history of $D_{eff}$ is indeed proportional to the ion heat conductivity $\chi_i$, which further verifies the diffusive nature of the ion transport.

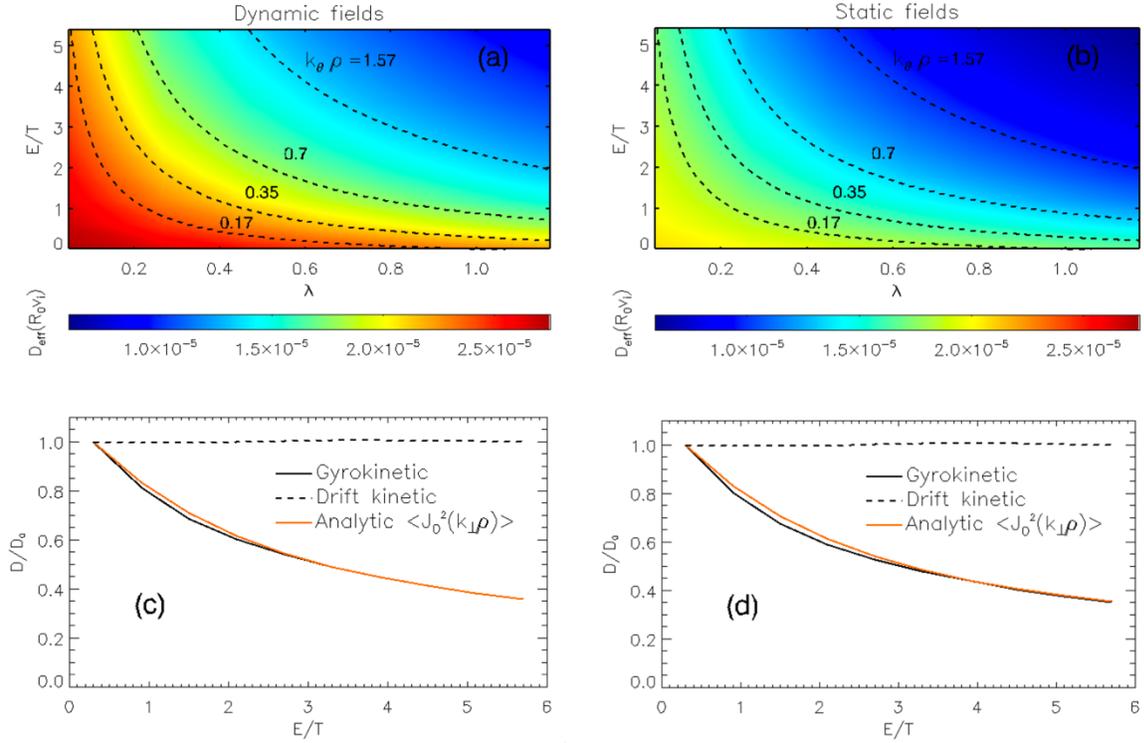

**Figure 2.** Phase space ($E$, $\lambda$) structure of ion mobility $D_{\text{eff}}$ in the ITG non-linear regime with self-consistent fields (a) and with fixed fields (b). The dashed black lines represent constant $k_\theta\rho$ values. The phase space cut for the deeply trapped particles are shown in solid black line in panels (c) and (d) in the self-consistent field and fixed field respectively, demonstrating good agreement with analytic results of the gyro-averaging (red line). Drift kinetic test particle mobility is shown in dashed line in panels (c) and (d).

Figure 2 show the phase space ($E$, $\lambda$) structure of the ion mobility averaged over $25\, R_0/v_i$ in the nonlinear regime of the ITG turbulence with self-consistent fields (panel a) and with fixed fields (panel b). Here, $\lambda=\mu B_0/E$ is a pitch-angle $\xi$ related variable, $\xi=v_\parallel/v$. We find that the constant $D_{\text{eff}}$ curves in the ($E$, $\lambda$) phase space fit very well to the constant-$k_\theta\rho$ curves for both the self-consistent fields and the fixed fields. This indicates that the breaking of the constants of motion in both cases is due to the *EXB* drifts, which exhibits the phase space structure only in the form of $k_\theta\rho$ through the gyro-averaging of the turbulence fields. To further verify that the structure of panels (a) and (b) comes from the gyro-averaging, we take a cut of the (E, $\lambda$) phase space for the deeply trapped particles with $\xi \subset [0, 0.1]$. Panels (c) and (d) shown that the ion mobility (black curves) fits almost perfectly with the analytic results of the gyro-averaging (red curves), i.e., the square of the Bessel function $<J_0^2(k_\perp\rho)>/<J_0^2(k_\perp\rho_0)>$ for both the self-consistent fields and the fixed fields. Here $k_\perp = \sqrt{k_\theta^2 + k_r^2}$ is the perpendicular wave number and $<...>$ is average over the poloidal field spectrum. $k_\theta$ and $k_r$ are chosen with fixed weight in this calculation according to the 1-D poloidal and toroidal spectrum shown in panel (a) of Figure 6. Since in ITG turbulence the average $k_r$ is much smaller than $k_\theta$, rendering the correction from $k_r$ to $k_\perp$ very small. In the CTEM turbulence discussed in the next section, $k_r/k_\theta \sim 0.6$, bringing a twenty percent correction to $k_\perp$. Furthermore, we measured the ion mobility by using test guiding centers without the gyro-averaging processes. These test guiding centers do not feed back to the turbulence fields, and therefore do not affect the turbulence dynamics. The ion mobility measured from the test guiding centers shown in dashed curves in panels (c) and (d) of Figure 2 is indeed almost uniform in both self-consistent fields and in fixed fields.

In summary, the breaking of the adiabatic invariants leading to the ion transport in the ITG turbulence is dominated by the nonlinear *EXB* scattering. The linear resonance due to the time variations of the turbulence fields is subdominant.

### III. CTEM turbulence

To study the CTEM turbulence, we use the plasma parameters described above but with the ion temperature gradient of $R_0/L_{Ti}=2.2$ and electron temperature gradient of $R/L_{Te} = 6.9$. The CTEM is the dominant instability for these parameters. The trapped electrons dynamics are simulated by the drift kinetic equation using the fluid-kinetic hybrid electron model [17] and the passing electrons are assumed to be adiabatic. The time evolution of the CTEM-induced electron heat transport, ion heat transport and particle transports from the simulation is shown in panel (a), (b) and (c) in Figure. 3 respectively. The self-consistent particle flux $\Gamma = \int dv^3 \, \delta v_{E \times B} \delta f$ is used to define the particle diffusivity D by using the relation $\Gamma = D\nabla n$. The CTEM instability initially grows exponentially in the linear regime. In the nonlinear regime, the electron heat flux saturates first as shown in Figure 3(a). The ion heat flux and particle flux then also saturate to reach a fully developed turbulence.

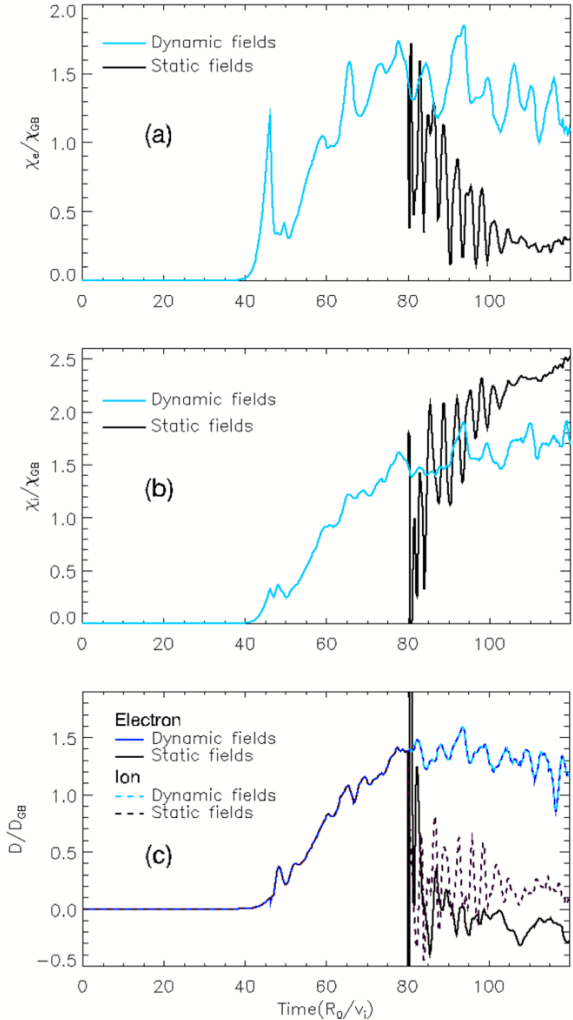

**Figure 3.** Time history of **(a)** electron heat conductivity $\chi_e$, **(b)** Ion heat conductivity $\chi_i$, and **(c)** particle diffusivity D with self-consistent fields (blue lines) and with fixed fields in the nonlinear regime (black

lines). Electron heat and particle transport drops nearly to zero when turbulence fields are fixed, while ion heat transport level changes only quantitatively.

In the wave-dominated turbulence, electron transport is induced by the breaking of the third adiabatic invariants mostly due to the time variations of the fluctuating fields (linear resonance). To verify the dominance of the linear resonance, we compare the transport between fully self-consistent simulation with dynamic fields and test-particle simulation with fixed turbulence fields at $t=80\ R_0/v_i$. We find that the electron heat conductivity and particle diffusivity drop nearly to zero with fixed turbulence fields, indicating the dominance of the linear resonance in driving the electron transport. On the other hand, the ion heat conductivity from the fixed turbulence fields changes only slightly compared with that from the self-consistent fields, suggesting that the *EXB* scattering dominate the ion nonlinear dynamics in CTEM turbulence. In Figure. 3, the ion and electron particle diffusivity are identical with the self-consistent turbulence fields as expected from the quasi-neutrality condition, but has a small difference in the residues with fixed turbulence fields, which can be caused by the breaking of quasi-neutrality condition since the gyrokinetic Poisson's equation is no longer fully solved when the non-zonal component of the electrostatic potential is fixed. To test the role of the self-consistent zonal flows, we perform the same simulations of the ITG and CTEM turbulence, but fixing both the zonal flow and the non-zonal components of the turbulence fields. The electron and ion heat transport evolve in similar trend without the GAM oscillation. In the CTEM turbulence, ion particle flux remains a finite value, and deviates dramatically from electron particle flux.

To further verify the importance of the linear resonance and nonlinear scattering in driving electron and ion transport, respectively, we study the phase space structures of ion and electron mobility. We continue to use $(E, \lambda)$ as the phase space coordinate for ions and electrons. For trapped electrons, $\lambda \subset [1-\varepsilon, 1+\varepsilon]$. In this diagnosis, ions and electrons are selected from around the $r=0.5a$, therefore $\varepsilon=r/R_0$ is approximately *0.18*, and $\lambda \subset [0.82, 1.18]$. For both species there are *10* bins in each dimension, and more than *5000* particles in each bin. For enhanced statistics all the values are averaged over a time period of $25R_0/V_i$.

In the CTEM turbulence, trapped electron motion is not purely diffusive, and the mobility calculated in Eq. 1 is no longer a good representation for the heat conductivity and particle diffusivity. Therefore we use the effective heat conductivity defined as $\chi_{eff} = q(E,\lambda)T/(n_0 \nabla T E)$ to represent the phase space distribution of the heat conductivity by using the heat flux at local phase space. The distribution of electron effective heat conductivity in the non-linear stage in Figure 4 (a) is consistent with previous CTEM simulation results using Lagrangian analysis [6], which verifies that $\chi_{eff}$ is a proper substitute for trapped electron transport.

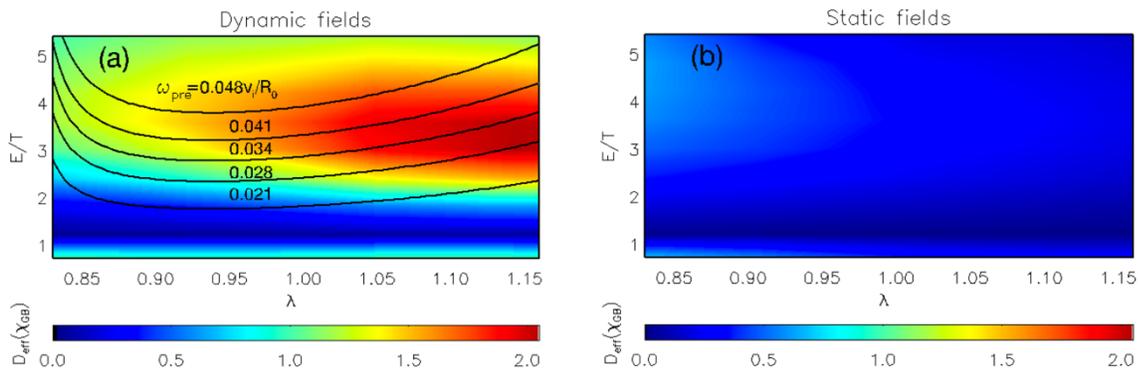

**Figure 4.** Electron mobility $\chi_{eff}$ in 2-D phase space (E,λ) **(a)** in the nonlinear stage with self-consistent fields and **(b)** with fixed fields. The black lines in **(a)** represent the electron precessional frequency.

Figure 4**(a)** indicates that before freezing the fluctuating fields, electron heat transport mainly comes from the contribution of deeply trapped particles around the energy $\sim 4T_e$. According to quasi-linear theory, the resonance condition for trapped particle is $\omega = n\omega_{pre} + p\omega_b$, in which ω is the real frequency of the field, $\omega_{pre}$ and $\omega_b$ are the precessional and bounce frequency, n is the toroidal mode number. Take $p=0$ in the delta function since electron bounce motion is much faster than the CTEM frequency, we can get the resonance condition $\omega_{pre}=\omega/n$. The linear CTEM frequency is nearly dispersiveless, $\omega_r^{lin} = 1.52 k_\theta \rho_i v_i / R_0$ for wavevector range of $0<k_\theta\rho_i<1$. The nonlinearly dominant mode is $k_\theta\rho_i=0.3$ with $n=27$ in the nonlinear regime. This gives $\omega/n =0.034 v_i/R_0$, which is consistent with the mobility distribution in Fig. 4**(a)**. After the fields are fixed, the distribution evolved to almost uniform with small value, as shown in Fig. 4**(b)**. This is also consistent with our observation in Figure 3. The structure of the ion mobility is shown in Fig. 5, which exhibits similar features in ITG turbulence.

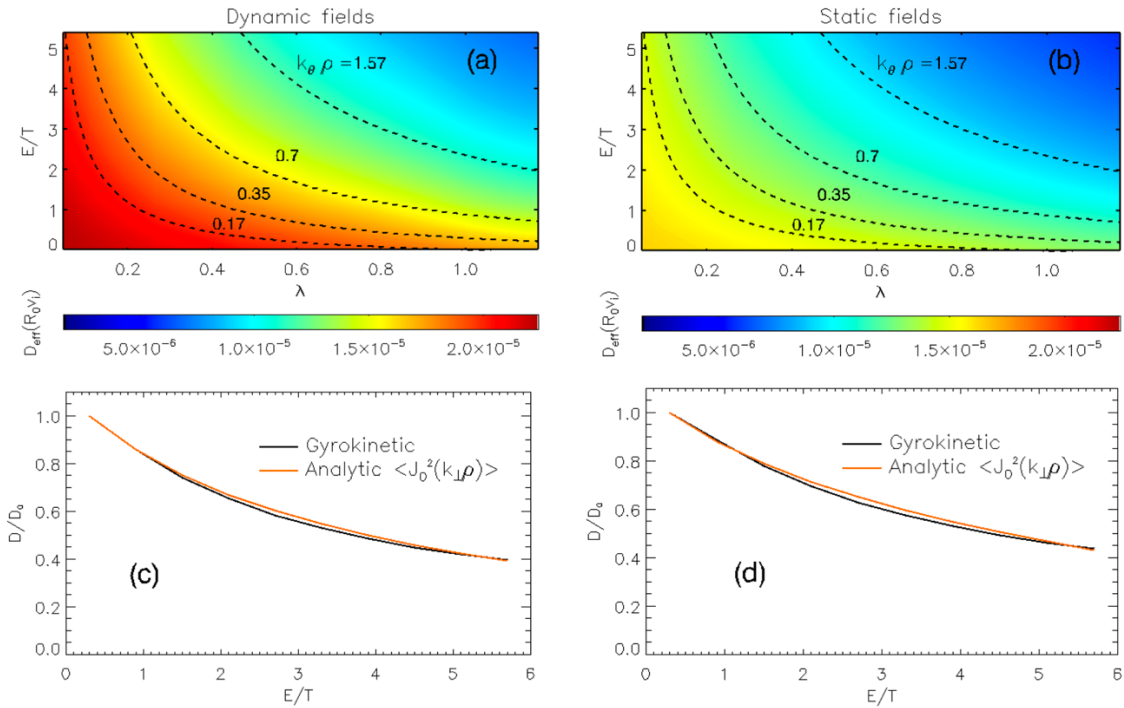

**Figure 5.** Ion mobility $D_{eff}$ in the non-linear stage in CTEM with self-consistent field **(a)** in 2-D phase space and **(c)** for the deeply trapped particles and with fixed field **(b)** in 2-D phase space and **(d)** for the deeply trapped particles. The black lines in both **(a)** and **(b)** represent fixed $k_\theta\rho$ values.

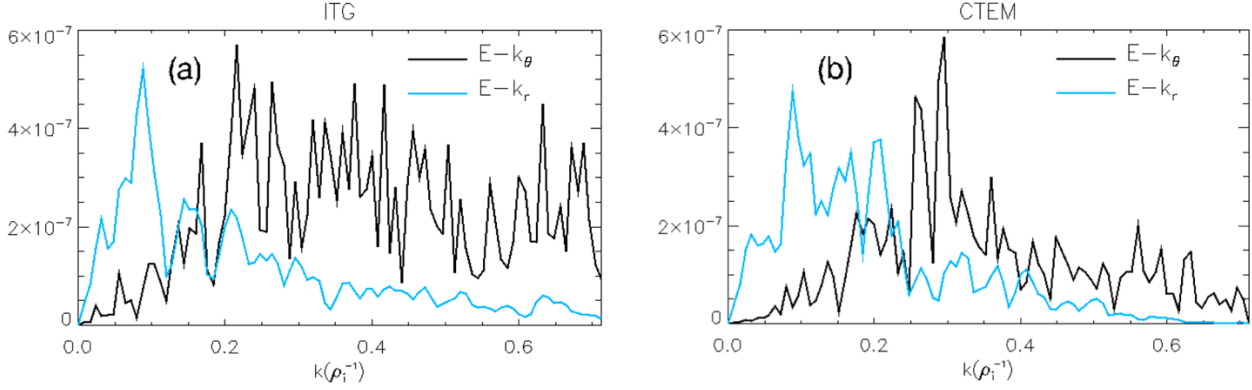

**Figure 6.** ITG **(a)** and CTEM **(b)** turbulence poloidal (black) and radial (blue) spectrum.

## IV. CONCLUSIONS

To summarize the results from CTEM and ITG simulations, electron transport is primarily regulated by wave-particle linear resonance, and the ion transport is regulated by nonlinear wave-particle decorrelation.

For trapped electrons, ($\mu$, $J_{//}$) are conserved, and the linear precessional resonance $\omega_{pre}=\omega$ breaks the third invariant S and induces radial transport since the non-linear decorrelation is weak [9]. After the field is fixed, $\omega=0$, the transport reduces greatly due to the removal of linear resonance. The nonlinear frequency associated with E×B drift might induce small transport by breaking S invariant.

The ion transport is a quasilinear process and regulated by wave-particle decorrelation with the parallel decorrelation time $\tau_{//}$ and the time scale of electron diffusion across the radial structure $\tau_{\perp}$, which is independent of time variation of $\delta\Phi$. Therefore, when the field is fixed, ion transport does not encounter much change.

This might be able to explain the strong linear resonance observed in the transport of trapped electrons, and weak linear resonance observed in the transport of ions. These findings cast further doubts on the validity of the quasilinear theory for modeling the particle and heat transport in the toroidal plasma micro-turbulence.

**ACKNOWLEDGMENTS:**
This work was supported by US Department of Energy (DOE) SciDAC GSEP Center (Grant No. DE-FC02-08ER54976) and China Scholarship Council (Grant No. 201206010268). Research at the Princeton Plasma Physics Laboratory is supported by the US DOE contract DE-AC02-09CH11466. This work used resources of the Oak Ridge Leadership Computing Facility at Oak Ridge National Laboratory (DOE Contract No. DE-AC05-00OR22725) and the National Energy Research Scientific Computing Center (DOE Contract No. DE-AC02-05CH11231).